\documentclass[aps,prb,showpacs,twocolumn]{revtex4}
\usepackage{epsfig,graphicx}
\usepackage{amssymb}
\begin{document}
\title{Application of screened hybrid functionals to the bulk transition
metals Rh, Pd, and Pt}
\author{Fabien Tran}
\author{David Koller}
\author{Peter Blaha}
\affiliation{Institute of Materials Chemistry, Vienna University of Technology,
Getreidemarkt 9/165-TC, A-1060 Vienna, Austria}

\begin{abstract}

We present the results of calculations on bulk transition metals Rh, Pd, and Pt
using the screened hybrid functional YS-PBE0
[F. Tran and P. Blaha, Phys. Rev. B \textbf{83}, 235118 (2011)].
The results for the equilibrium geometry are compared with
those obtained from (semi)local functionals, namely, the local density
approximation and the generalized gradient approximation
PBE of Perdew \textit{et al}. [J. P. Perdew, K. Burke, and M. Ernzerhof,
Phys. Rev. Lett. \textbf{77}, 3865 (1996)]. It is shown that the screened hybrid
functional yields more accurate equilibrium geometry than PBE, but,
overall, it is not more accurate than LDA.
However, in contradiction with experiment, we find that
the screened hybrid functional favors a ferromagnetic state as the
ground state for all three transition metals. Therefore, the use of
hybrid functionals for, e.g., the study of
catalytically active systems with correlated
oxides on a metal support is questionable.

\end{abstract}

\maketitle

\section{Introduction}

The Kohn-Sham (KS) version of density functional theory (DFT) is the most used
quantum method for the calculations of properties of matter,
\cite{HohenbergPR64,KohnPR65} however, since we do not know the exact
exchange-correlation functional, one needs to make a choice for an approximate
form for practical calculations (see Ref. \onlinecite{CohenCR11} for a recent
review). Many different types of approximate exchange-correlation functionals
have been proposed. The two most commonly used approximations in solid-state
physics are the local density approximation (LDA)\cite{KohnPR65} and
generalized gradient approximation (GGA), which lead, in most cases, to
reasonably accurate results for the structural parameters in particular
(see, e.g., Refs. \onlinecite{HaasPRB09} and
\onlinecite{CsonkaPRB09}).

However, a well-known problem of LDA and GGA, is their inability to yield
accurate excited-states properties, but actually this is a more general problem
which has its roots in the KS method itself when an orbital-independent
potential is used (see Ref. \onlinecite{KummelRMP08}). The most common way
to get more accurate excited-state properties is to
work in the framework of the generalized KS scheme \cite{SeidlPRB96} by
using functionals which lead to orbital-dependent potentials like
DFT+$U$\cite{AnisimovPRB91} or the hybrids.\cite{BeckeJCP93}
DFT+$U$ is computationally as cheap as a LDA/GGA, but can be applied only to
well localized strongly correlated electrons (typically $3d$ or $4f$ electrons).
The hybrid functionals, which mix semilocal (i.e., LDA or GGA) and Hartree-Fock
(HF) exchange, can be used more widely, however for solids, they lead to
calculations which are one or two orders of magnitude more expensive than
LDA/GGA and their applications to metals can be problematic.\cite{GerberJCP07}

A way to reduce partially the problems encountered with hybrid functionals for
solids is to get rid of the long-range HF interaction by screening the Coulomb
potential with, e.g., an exponential \cite{BylanderPRB90} or error function.
\cite{HeydJCP03} Actually, screened hybrid functionals are becoming more and more
popular and have shown to be successful for structural and electronic properties
on a large variety of solids (see Refs.
\onlinecite{HeydJCP05,PaierJCP06,ClarkPRB10,ClarkPSSB11,HendersonPSSB11,MarquesPRB11,SchimkaJCP11,PeveratiJCP12}
for extensive tests) including difficult cases like transition-metal (TM) oxides
and nitrides, where band gaps and magnetism are strongly enlarged in agreement
with experiment (see, e.g., Refs.
\onlinecite{KudinPRL02,ProdanJCP05,ProdanPRB06,MarsmanJPCM08,EyertPRL11,SchlipfPRB11,BotanaPRB12}).
In the case of a GGA-based screened hybrid functional, the functional reads
\begin{equation}
E_{\text{xc}} = E_{\text{xc}}^{\text{GGA}} +
\alpha_{\text{x}}\left(E_{\text{x}}^{\text{SR-HF}} -
E_{\text{x}}^{\text{SR-GGA}}\right),
\label{Exchybrid}
\end{equation}
where $E_{\text{x}}^{\text{SR-HF}}$ and $E_{\text{x}}^{\text{SR-GGA}}$ are the
short-range (SR) parts of the HF and GGA exchange functionals, respectively,
and $\alpha_{\text{x}}$ ($\in [0, 1]$) is the fraction of exact exchange.
The most known screened hybrid functional is the one proposed by Heyd
\textit{et al}. (HSE),\cite{HeydJCP03} which is based on the GGA functional
PBE of Perdew \textit{et al}.\cite{PerdewPRL96} (like the unscreened hybrid
functional PBE0\cite{ErnzerhofJCP99,AdamoJCP99}) and uses the error function
for the splitting of the Coulomb potential. Most of the time, HSE is used with
$\alpha_{\text{x}}=0.25$. Another known screened hybrid functional is the early one
proposed by Bylander and Kleinman,\cite{BylanderPRB90} which is based on the LDA
functional and uses a screening parameter that is calculated from the average value
of the valence electron density (in HSE the screening parameter is fixed
to a ``universal'' value).

The trends of the HSE06\cite{KrukauJCP06} functional is to be very accurate
for band gaps smaller than $\sim5$ eV, but to underestimate larger band gaps.
Actually, the larger the band gap, the more the underestimation
(in percentage) becomes severe (see Refs. \onlinecite{MarquesPRB11} and
\onlinecite{BrothersJCP08}).
Usually, GGA functionals underestimate the atomic magnetic moments in
strongly correlated systems, while the values with HSE06 (or any other
hybrid functional) are increased and hence in better agreement with experiment.
\cite{MarsmanJPCM08} Concerning the lattice constants,
\cite{SchimkaJCP11,PeveratiJCP12} HSE06 is more accurate than PBE, but not as
accurate as the GGA functionals which were especially designed for solids like
WC\cite{WuPRB06} or PBEsol.\cite{PerdewPRL08} However, we note that the values
of the lattice constants depend strongly on the GGA functional on which the
hybrid functional is based, while the band gaps are rather insensitive
(see Ref. \onlinecite{SchimkaJCP11}).
Actually, recent studies reported calculations on solids using hybrid functionals
based either on the WC\cite{BilcPRB08} or PBEsol\cite{SchimkaJCP11}
GGA functional, and in both cases it was shown that the hybrid
functional benefits from the good performances of the
underlying GGA (WC or PBEsol) for geometrical parameters.
We also mention that in
Ref. \onlinecite{LuceroJPCM12}, a middle-range hybrid functional 
was shown to perform very well on semiconductors for the lattice constant and
band gap. Recently (Ref. \onlinecite{TranPRB11}),
a screened hybrid functional based on the GGA PBE functional, but using the
exponential function for the splitting of the Coulomb potential (known as the
Yukawa potential), was implemented into the WIEN2k code.\cite{WIEN2k}
The choice of using the exponential function instead of the error function
was made for technical reasons (e.g., integrals calculated more easily),
but is of very little importance for applications.
Indeed, it was shown that using the screening parameter $\lambda=0.165$ bohr$^{-1}$
($\lambda$ was kept fixed at this value for the present work)
in the Yukawa potential $e^{-\lambda\left\vert\mathbf{r}-\mathbf{r}'\right\vert}
/\left\vert\mathbf{r}-\mathbf{r}'\right\vert$
leads to results which are close to the results obtained with HSE06
\cite{KrukauJCP06} for which $\mu=0.11$ bohr$^{-1}$ in the SR potential
$\text{erfc}\left(\mu\left\vert\mathbf{r}-\mathbf{r}'\right\vert\right)
/\left\vert\mathbf{r}-\mathbf{r}'\right\vert.$
The functional was called YS-PBE0 (where YS stands for Yukawa screened).

In this work, we present the results of calculations obtained with the
screened hybrid functional YS-PBE0 on the face-centered cubic $4d$-TM
Rh and Pd and $5d$-TM Pt. Usually, the semilocal functionals
describe fairly well the electronic and magnetic properties of the
(non)-magnetic $3d$, $4d$, and $5d$ TM, although in particular for GGA
functionals a tendency to overestimate magnetism can been found.
\cite{SinghPRB92,MazinPRB04,AguayoPRL04,SubediPRB10,OrtenziPRB12}
This is in contrast to the TM oxides (in particular the $3d$ like MnO,
NiO,\cite{TerakuraPRB84} or the undoped cuprates\cite{PickettRMP89}),
where magnetism is underestimated and semilocal functionals
give qualitatively wrong results due to the localized nature of the
$d$-electrons, while in the pure TM the electrons are itinerant
(i.e., more free-electron like).

\section{results}

\begin{table}
\caption{\label{table1}Equilibrium lattice constant $a_{0}$ (in \AA),
bulk modulus $B_{0}$ (in GPa), and the total-energy difference
(in mRy/atom) between the minima of the FM and NM phases. The experimental
values are for $T=0$ K and are corrected for the zero-point anharmonic
expansion.\cite{Lejaeghere}}
\begin{tabular}{lccccc}
\hline
\hline
Method & $a_{0}^{\text{NM}}$ & $a_{0}^{\text{FM}}$ &
$B_{0}^{\text{NM}}$ & $B_{0}^{\text{FM}}$ &
$E_{\text{tot}}^{\text{NM}}-E_{\text{tot}}^{\text{FM}}$ \\
\hline
\multicolumn{6}{c}{Rh} \\
LDA     & 3.757 &       & 316 &     &      \\
PBE     & 3.831 &       & 257 &     &      \\
YS-PBE0 & 3.799 & 3.874 & 279 & 226 & 0.59 \\
Expt.   & 3.786 &       & 277 &     &      \\
\hline
\multicolumn{6}{c}{Pd} \\
LDA     & 3.840 &       & 228 &     &      \\
PBE     & 3.943 & 3.946 & 168 & 165 & 0.15 \\
YS-PBE0 & 3.935 & 3.938 & 165 & 168 & 3.96 \\
Expt.   & 3.876 &       & 187 &     &      \\
\hline
\multicolumn{6}{c}{Pt} \\
LDA     & 3.896 &       & 305 &     &      \\
PBE     & 3.971 &       & 246 &     &      \\
YS-PBE0 & 3.948 & 3.959 & 265 & 261 & 3.51 \\
Expt.   & 3.917 &       & 286 &     &      \\
\hline
\hline
\end{tabular} 
\end{table}

\begin{figure}
\includegraphics[scale=0.7]{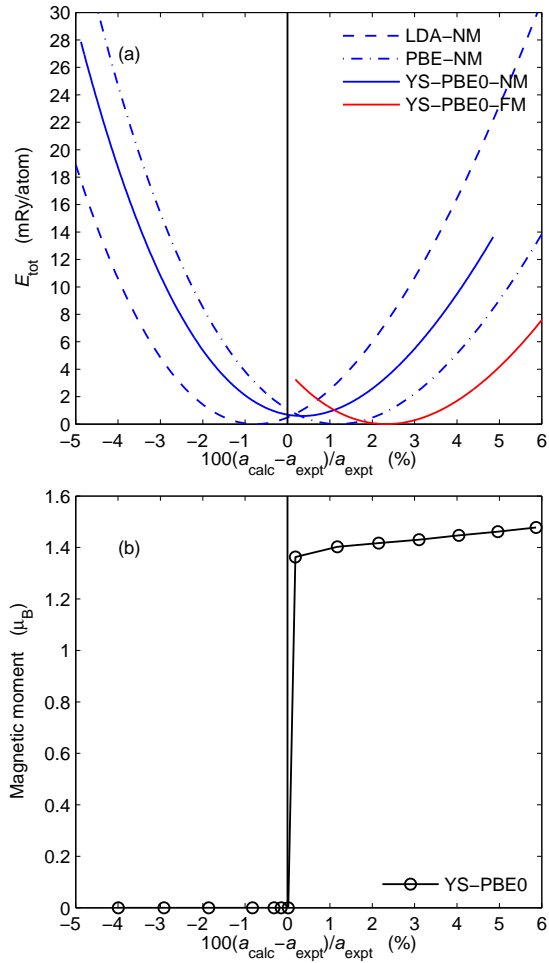}
\caption{\label{fig1}(Color online)
(a) Total energy and (b) magnetic moment in the unit cell of the FM and NM
phases of Rh versus the lattice constant (expressed as the relative difference
with respect to the experimental value) calculated from several functionals.
For each functional, the zero of the energy was set to the lowest minimum.
The vertical line indicates the experimental lattice constant.}
\end{figure}

\begin{figure}
\includegraphics[scale=0.7]{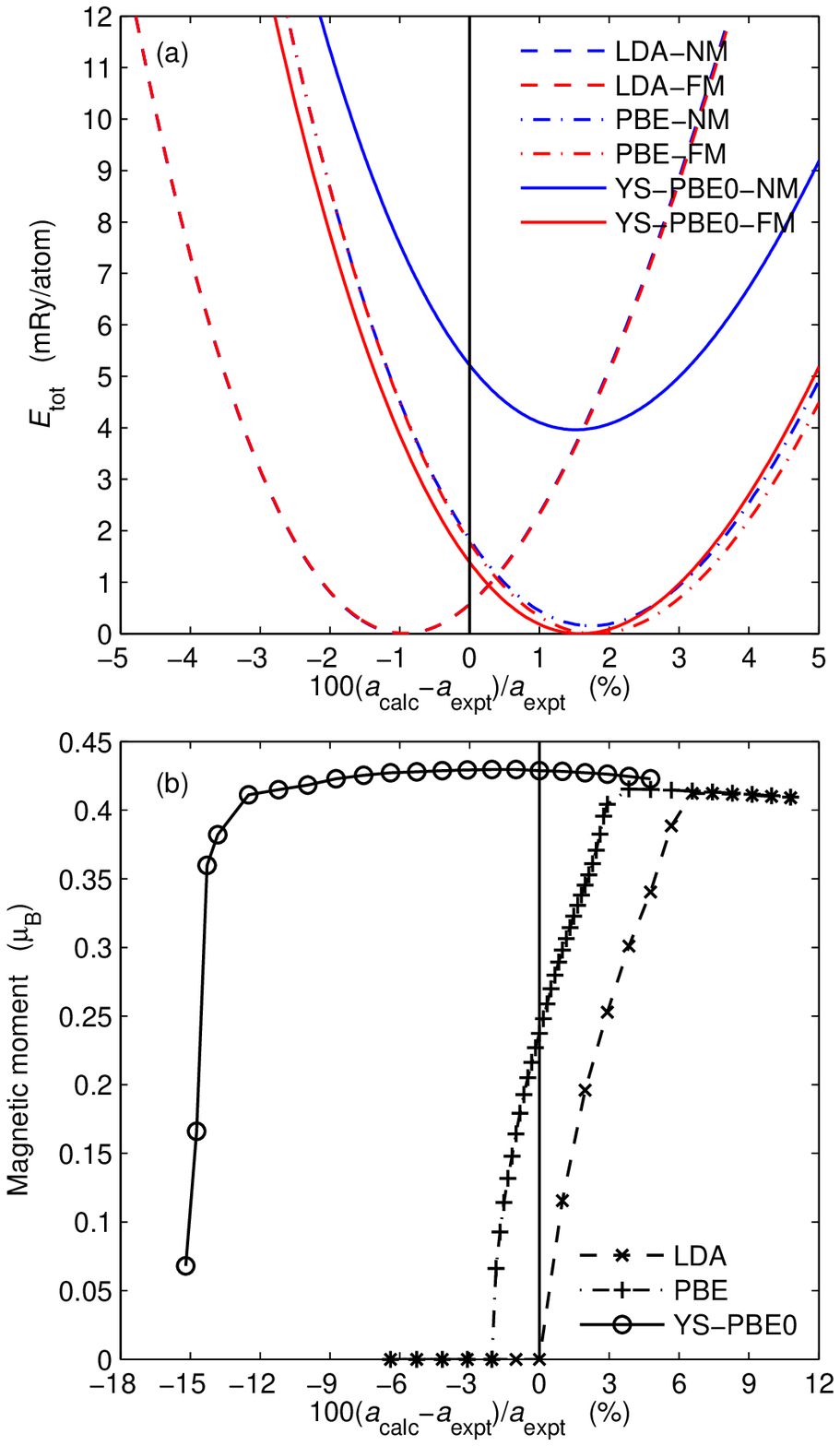}
\caption{\label{fig2}(Color online)
Same as Fig. \ref{fig1} but for Pd. Note that in (a) and (b) the ranges for
the $x$ axis are different.}
\end{figure}

\begin{figure}
\includegraphics[scale=0.7]{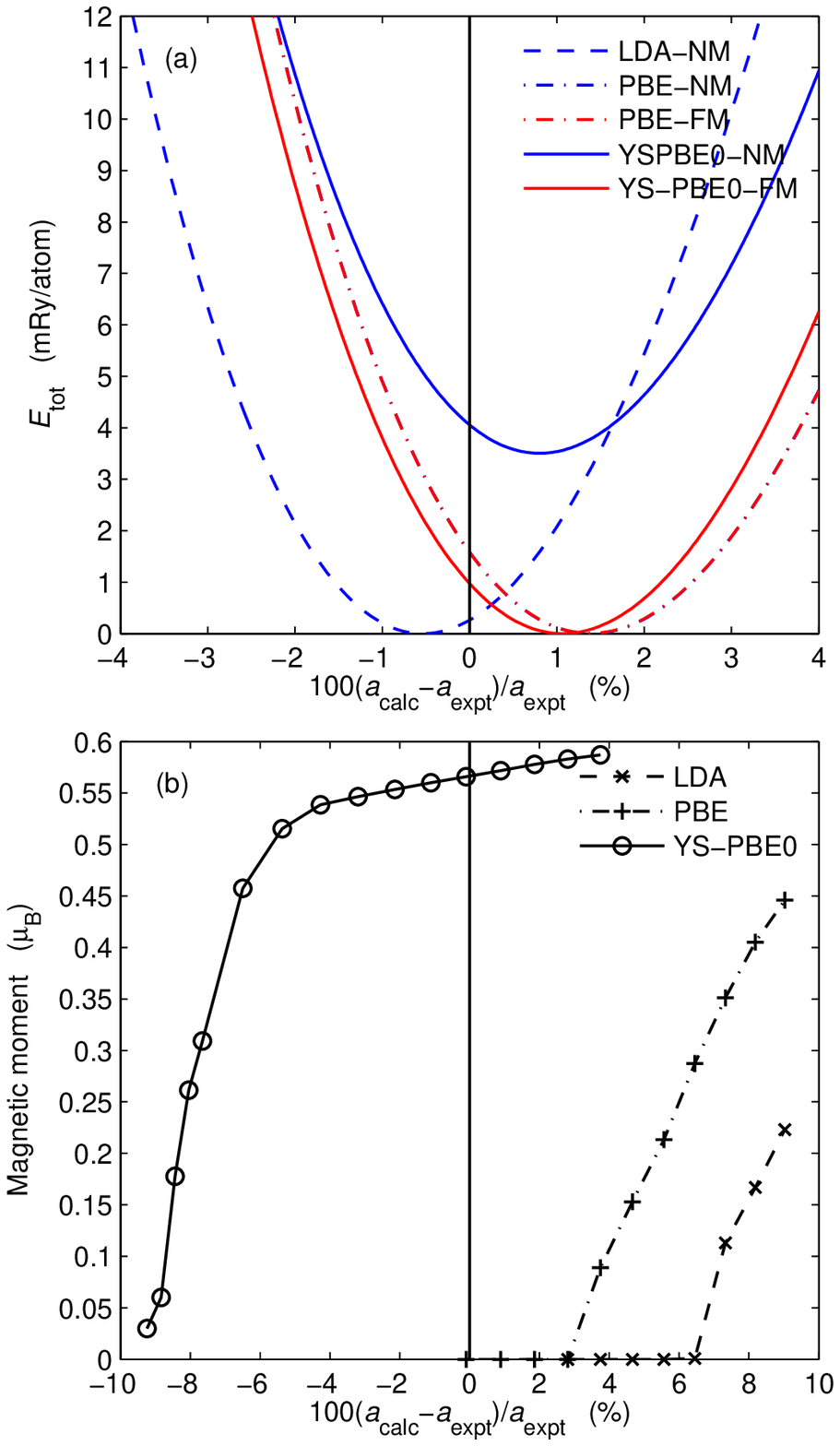}
\caption{\label{fig3}(Color online)
Same as Fig. \ref{fig1} but for Pt. Note that in (a) and (b) the ranges for
the $x$ axis are different.}
\end{figure}

The calculations were done with the WIEN2k code,\cite{WIEN2k}
which is based on the full-potential (linearized)
augmented plane-wave and local orbitals method to
solve the KS equations.\cite{Singh} For comparison purposes,
calculations with the semilocal functionals LDA
(using PW92\cite{PerdewPRB92} for correlation) and the GGA
PBE\cite{PerdewPRL96} were also done. The semilocal calculations were done
with a $24\times24\times24$ $\mathbf{k}$-mesh
for the integrations of the Brillouin zone, while a
$12\times12\times12$ $\mathbf{k}$-mesh was used for the hybrid functional
YS-PBE0 to make the computational time reasonable. We checked that using
such a $\mathbf{k}$-mesh leads to results which are
accurate enough for our purposes.
The size of the basis set is determined by the
product $R_{\text{MT}}^{\text{min}}K_{\text{max}}$ of the smallest of
the atomic sphere radii ($R_{\text{MT}}$)
and the plane wave cutoff parameter ($K_{\text{max}}$).
The values 9 (100-180 basis functions per atom) and 10
(160-280 basis functions per atom) for
$R_{\text{MT}}^{\text{min}}K_{\text{max}}$ were
used for the hybrid and semilocal calculations, respectively.
The atomic sphere radius
$R_{\text{MT}}$ was chosen as 2.2 bohr for Rh and 2.1 bohr for Pd and Pt.

The results of the calculations obtained with the LDA, PBE and YS-PBE0
functionals are shown in Table \ref{table1} and Figs. \ref{fig1}-\ref{fig3}.
First, we start the discussion about the equilibrium geometry of the
experimentally observed non-magnetic (NM) phase.
The experimental values for $a_{0}$ (taken from Ref. \onlinecite{Lejaeghere})
were extrapolated to $T=0$ K (using the volume expansion
coefficient) and corrected for the zero-point
oscillations according to the scheme of Alchagirov
\textit{et al}.\cite{AlchagirovPRB01}
For Rh, we can see that the hybrid functional YS-PBE0
gives a fairly accurate value of 3.799 \AA~(3.786 \AA~for experiment),
while LDA underestimates $a_{0}$ by 0.03 \AA~and PBE overestimates
by 0.045 \AA. In the case of Pd, LDA, which leads to an underestimation of
$\sim0.035$ \AA, is more accurate than PBE and YS-PBE0 which lead to rather
large overestimations ($>0.06$ \AA).
For the $5d$-TM Pt, LDA is more accurate (underestimation of only
$\sim0.02$ \AA) than PBE and YS-PBE0 which yield too large equilibrium
lattice constant. By comparing our values for $a_{0}$ to the PBE and HSE06
results from Ref. \onlinecite{SchimkaJCP11} for Rh and Pd, we can see that
the agreement is perfect (as expected) with PBE (less than 0.001 \AA~of difference),
while our YS-PBE0 lattice constants are about 0.015 \AA~larger than with
HSE06. As explained in Ref. \onlinecite{TranPRB11}, the differences
between the YS-PBE0 and HSE06 lattice constants are mainly due to the
different schemes used to screen the semilocal exchange energy.
In YS-PBE0, the simple scheme of Iikura \textit{et al}. \cite{IikuraJCP01}
was used, while
in HSE06 the screening is done by considering the expression
of the functional in terms of the exchange hole.\cite{ErnzerhofJCP98}

At this point it is useful to recall that the trends observed above with the
semilocal functionals are actually rather general among the elemental TM
(see, e.g., Ref. \onlinecite{HaasPRB09}). For the $3d$ series, PBE is the
most accurate, while LDA strongly underestimates
the lattice constants. For the $4d$ TM, GGA functionals which are softer
than PBE (e.g., WC\cite{WuPRB06} or PBEsol\cite{PerdewPRL08}) are more
accurate than LDA and PBE, while for the $5d$, LDA is the most accurate method.
It has also been observed (see, e.g., Ref. \onlinecite{SchimkaJCP11}) that
for most solids (Li and Na are two exceptions), replacing a fraction of
semilocal exchange by HF exchange leads to a reduction of the equilibrium
lattice constant, which is what has also been observed in the present work
with Rh, Pd, and Pt (i.e., the YS-PBE0 lattice constants are smaller than the
PBE ones). Therefore, the hybrid functionals based on PBE (HSE and YS-PBE0)
correct the PBE overestimation for the lattice constants.
\cite{HeydJCP05,SchimkaJCP11} This also explains why the lattice constants
calculated with the screened-exchange LDA\cite{BylanderPRB90}
(100\% of LDA exchange replaced by HF exchange)
are in most cases by far too small,\cite{ClarkPRB10} 
since on average LDA already underestimates them.
The results for the bulk modulus $B_{0}$ (see Table \ref{table1}) show the usual
trends: An underestimation (overestimation) of $B_{0}$ is associated to an
overestimation (underestimation) of $a_{0}$.

Spin-polarized calculations on Rh, Pd, and Pt were also done in attempts to
stabilize a ferromagnetic (FM) solution. The results, whenever available, are
shown in Table \ref{table1} and Figs. \ref{fig1}-\ref{fig3}. In the case of Rh,
it was not possible to stabilize a solution with a non-zero magnetic moment
with LDA for the range of lattice constants that we considered (up to 15\%
larger than $a_{\text{expt}}$), while with PBE it is for values of $a$ larger
than $a_{\text{expt}}$ by about 13\% (i.e., much larger than the equilibrium
lattice constant) that such a solution could be obtained. With YS-PBE0 it was
possible to stabilize a magnetic moment for lattice constants larger than the
experimental one. Interestingly, the minimum of the FM total-energy curve is
lower than the minimum of the NM curve by 0.59 mRy/atom [see Fig. \ref{fig1}(a)
and Table \ref{table1}]. This can be considered as a
failure of the screened hybrid functional which describes Rh as magnetic, and
in addition the equilibrium lattice constant is more than 2\% larger than in
experiment. The value of the magnetic
moment in the unit cell is around 1.4 $\mu_{\text{B}}$ [see Fig. \ref{fig1}(b)]
for lattice constants larger than $a_{\text{expt}}$ and drops suddenly to zero
(NM solution) at this geometry.

Turning now to Pd, which is known experimentally and theoretically to be on
the verge of becoming FM,\cite{OswaldPRL86} we were able to stabilize a FM
solution with all considered functionals. From Fig. \ref{fig2}(b), we can see
that the smallest lattice constant at which it was possible to obtain a FM
solution is around $-15$\%, $-1.7$\% and 1\% of $a_{\text{expt}}$ for
YS-PBE0, PBE, and LDA, respectively, and therefore, at both the experimental
geometry as well as at the corresponding theoretical geometry, the magnetic
moment is non-zero for PBE and YS-PBE0. Note that the values of the magnetic
moments obtained at larger values of $a$ with the different functionals are
quite similar (0.40$-$0.45 $\mu_{\text{B}}$).
As for Rh with YS-PBE0, the FM minimum is lower than the NM minimum for PBE and
YS-PBE0 functionals by 0.15 and 3.96 mRy/atom, respectively. The latter value
is one order of magnitude larger than in the case of Rh, and
again this can be considered as a quite severe failure of the screened
hybrid functional.

For Pt [see Fig. \ref{fig3}(b)], non-zero magnetic moments can be obtained for
lattice constants larger than approximately $-9$\%, 3\%, and 7\% of
$a_{\text{expt}}$ for YS-PBE0, PBE, and LDA, respectively, and thus at the
experimental or theoretical lattice constant only YS-PBE0 gives a wrong
magnetic ground state. Thus, from Fig. \ref{fig3}(a) we can see again,
that with YS-PBE0 the FM state is more stable than the NM state and their
minima differ substantially by 3.51 mRy/atom.

\section{summary}

From the results presented above, we can conclude that the screened
hybrid functional YS-PBE0 slightly improves over PBE for the equilibrium
lattice constants of solids, which were too large with PBE (a rather general
trend with PBE). But note that
for the $4d$-TM Pd and $5d$-TM Pt, LDA is more accurate than PBE and YS-PBE0.
However, the main result of the present work is the inability of YS-PBE0
to give the correct ground state for the three considered TM. Indeed, the
FM state is found to be lower in energy than the experimentally observed
NM state. It is well-known (see Ref. \onlinecite{GerberJCP07}
and references therein) that the use of the HF approximation for metals
leads to severe problems like a vanishing density of states at the Fermi
level or too large magnetic moments.
(A remedy would be to use a compatible correlation functional
like the random-phase approximation,\cite{KotaniJMMM98} but such methods
are very expensive and their
self-consistent implementation not trivial at all.)

These problems are in particular due to the long-range part of the HF
exchange, which is not present in screened hybrid functionals.
However, it was shown\cite{PaierJCP06} that the screened hybrid functionals
still lead to an overestimation of the exchange splitting (and the
corresponding magnetic moment)
in itinerant magnetic systems like Fe for which semilocal methods work
quite well.

While the use of the gradient correction was necessary to get correctly
the body-centered cubic FM phase as the ground
state in Fe\cite{BagnoPRB89,BarbielliniJPCM90} (the ground state is
face-centered cubic NM with LDA), our
work has shown more evidence that using screened exact exchange as an additional
ingredient in functionals is not recommended for itinerant metals.
This has important implications for the studies of correlated TM oxides on TM
surfaces, which is relevant for many important catalytic active surfaces,
where oxides on metals like Rh, Pd, or Pt play a crucial role.
While the hybrid functionals describe in such systems the correlated oxides
pretty well, the results suffer from the overestimation of magnetism for the
underlying metals\cite{FranchiniJCP09} and thus the DFT+$U$ methods, where the
strength of the correlation effects
unfortunately has to be selected by hand for each atom, seem to be the only
practical way at the moment to describe such systems.

\begin{acknowledgments}

This work was supported by the project SFB-F41 (ViCoM) of the Austrian
Science Fund.

\end{acknowledgments}

\end{document}